# Imaging Spectrophotometry of Markarian 573

Richard W. Pogge[1]

Department of Astronomy, The Ohio State University, Columbus, Ohio 43210, USA

email: pogge@payne.mps.ohio-state.edu

Michael M. De Robertis[1]

Dept. of Physics & Astronomy, York University, North York, Ontario M3J 1P3, Canada

and

European Southern Observatory, Karl-Schwarzschildstraße 2, D85748 Garching bei München, Germany

email: mmdr@lhotse.phys.yorku.ca

## ABSTRACT

Narrow-band emission-line and continuum images obtained in subarcsecond seeing conditions are presented for the Seyfert 2 galaxy, Markarian 573. Images in the emission-lines of [O I], [O II], [O III], [S II], and H$\alpha$+[N II] are used to measure the structure and excitation in the extended emission regions surrounding the active nucleus. The biconical region contains a variety of complex features, including two arc-like structures within $\sim 2''$ of the active nucleus that appear to be associated with the extended radio continuum emission in the sense that the radio lobes lie just inside the centroids of the emission-line arcs. These arcs probably arise from an interaction between the ISM and the radio plasma from the active nucleus. Two other emission-line features lie $\sim 3-4''$ on either side of the nucleus, and appear to be elongated nearly perpendicular to the symmetry axis of the inner radio and optical structures. The existence of multiple emission-line components could be evidence for episodic ejection of radio plasma from the active nucleus.

*Subject headings:* galaxies:nuclei—galaxies:Seyfert—galaxies:active—galaxies:individual:Markarian 573

## 1. Introduction

Extensive regions of high surface-brightness, high excitation emission-line gas are found in the inner few kiloparsecs surrounding many active galactic nuclei (AGN) of the Seyfert class. In a number of Seyferts, these extended emission-line regions manifest a high-degree of axisymmetry, the most visually impressive of which are the "ionization cones" that appear to emanate from the active nucleus (e.g., Baldwin, Wilson, & Whittle 1987; Pogge 1989; Hutchings & Neff 1989; Haniff, Ward, & Wilson 1991; Storchi-Bergmann, Wilson & Baldwin 1992; Wilson et al. 1993). The emission-line gas in the cone regions have spectra consistent with photoionization by a hard-photon spectrum (e.g., Pogge 1988b; Tsvetanov & Walsh 1992, hereafter TW92). The strong symmetry of these regions is often taken as evidence that the photoionizing continuum radiation originating in the active nucleus is emerging along a preferred direction. Numerous comparisons between the optical and radio structure in these galaxies (using both long-slit and imaging data) have shown that this preferred direction is most clearly defined by the linear radio structure axis (Unger et al. 1987; Whittle et al. 1988; Haniff, Wilson, & Ward 1988; Bower et al. 1994; Wilson & Tsvetanov 1994). Beyond a shared orientation in projection, a few of the extended emission-line systems have detailed structures suggestive of a mechanical interaction between the radio continuum plasma and the optical line-emitting clouds in the form of direct spatial coincidence (e.g., Bower et al. 1994), structures and/or kinematic signatures suggestive of large-scale shocks (e.g., Wilson et al. 1985; Wilson & Ulvestad 1987 hereafter WU87; Whittle et al. 1988; Cecil, Bland & Tully 1990), or apparent affinities between the emission-line regions and the properties of the host galaxies (Whittle 1992a,b; Wilson & Tsvetanov 1994).

One of the best objects in which to study the interaction between an active nucleus and its host galaxy environment is the Seyfert 2 galaxy Markarian 573 (Koski 1978). Mrk 573 has a bright extended emission region (Unger et al. 1987; Haniff, Wilson, & Ward 1988), a linear radio continuum source (Ulvestad & Wilson 1984; henceforth UW84), strong, high-ionization emission-lines (TW92), and kinematical substructure in the [O III] emission line suggestive of gas disturbed in a jet/ISM interaction (Whittle et al. 1988; TW92). Moreover, neither the featureless optical continuum nor the emission-line radiation from the central source dominates emission from the immediate circumnuclear environment (e.g., De Robertis & Osterbrock 1986), as happens in so many AGNs. The host galaxy is a nearly face-on, early-type galaxy (morphological class RSAB(rs)0$^{+}$; de Vaucouleurs et al. 1991, RC3) with a heliocentric radial velocity of $5156 \pm 90$ km s$^{-1}$ (Whittle et al. 1988), giving a projected spatial scale in which 1″ corresponds to 0.33 $h_{75}^{-1}$ kpc. The galaxy itself does not appear to have any nearby companions (Dahari 1985), and its nearly face-on inclination ensures a fairly direct, unattenuated view of the nucleus and its surroundings.

The radio continuum maps of UW84 reveal a linear triple source consisting of the active nucleus and two lobes arranged symmetrically on either side of the nucleus. The lobes are separated from the nucleus by $\sim 1''\!.5$ and lie along a position angle of $122 \pm 3°$. Only the SE radio component is marginally resolved at this resolution, although they all have roughly the same integrated 6 cm radio fluxes. The kinematic substructure observed by Whittle et al. (1988) is in the [O III]$\lambda$5007 emission-line gas roughly coincident with the location of the off-nuclear radio lobes. The narrow-band H$\alpha$+[N II] and [O III] emission-line images of Haniff, Wilson, & Ward (1988) revealed that the extended "linear" distribution of high-excitation gas extends to much larger projected radii than the radio continuum lobes. Ionization maps of the optical emission-line regions show the familiar biconical morphology, although somewhat less prominently than the more classical examples of this phenomenon (e.g., NGC 4388, Pogge 1988b; NGC 5252, Tadhunter & Tsvetanov 1989). A detailed follow-up study combining imaging and long-slit spectroscopy (including kinematics and emission-line diagnostics at six different slit positions in the nuclear regions) by TW92 described physical conditions — in particular, density and excitation — as a function of position in the extended emission-line regions, and confirmed the interpretation of the [O III] observations made by Whittle et al. (1988), and Durret & Warin (1990) at a spatial resolution limited by seeing to $\sim 1''\!.3$. Because of the small angular separation of the radio lobes from the nucleus ($\sim 1''\!.5$), a detailed spatial comparison of the radio and optical properties has not been previously possible because of the substantial (nearly factor of 2) mismatch in spatial resolutions.

Conventional investigations of extended emission-line gas in galactic nuclei have concentrated on two wavebands: H$\alpha$+[N II]$\lambda\lambda$6548+6583 and [O III]$\lambda$5007 (e.g., Haniff, Wilson, & Ward 1988; Pogge 1989), with the occasional addition of a third filter band (e.g., [S II]$\lambda\lambda$6716+6731, TW92). A common technique employed in these studies is the construction of an "ionization map" by forming an image of the [O III]/(H$\alpha$+[N II]) line ratio (e.g., Pogge 1988a). This provides a semi-quantitative map of the ionization state of the gas, and may be used to separate (at least to first order) high-ionization gas excited by the active nucleus from lower ionization circumnuclear star formation regions. In many cases these maps have greatly facilitated the discovery of the ionization cones that are often characteristic of such systems. In this work, we extend the emission-line imaging technique to include five emission-lines: the lines of the three primary ionic species of oxygen accessible at visual wavelengths, [O I]$\lambda$6300, [O II]$\lambda$3727, and [O III]$\lambda$5007; the H$\alpha$+[N II]$\lambda\lambda$6548+6583 blend; and the [S II]$\lambda\lambda$6716+6731 doublet blend. With careful flux calibration, these emission-line images may be used to construct line-ratio maps that will provide more quantitative diagnostics of the ionization state than the traditional [O III]/(H$\alpha$+[N II]) ionization mapping technique.

In this paper we present new ground-based emission-line and continuum imaging for Mrk 573. Excellent (0″.60 ±0″.05 FWHM) seeing allows us to compare radio and optical properties at the same spatial resolution. These images reveal a great deal more structure in the extended emission-line regions than has been seen in previously published data. Our direct flux and line-flux ratio maps of the extended ionized gas regions provide compelling morphological and spectroscopic evidence for the strong interaction between the radio lobes and emission-line gas suggested by the earlier kinematical work. In particular, we see arc-like emission-line structures bracketing the radio lobes that are strongly reminiscent of hydrodynamic bowshocks (e.g., as seen in some Herbig-Haro objects such as HH-34, Morse et al. 1992). The observations and reduction techniques are presented in §2, and a description of the results are given in §3. Finally, §4 is devoted to a discussion of the results and our conclusions.

## 2. Observations & Data Reduction

We obtained direct emission-line and continuum-band images of Mrk 573 on the nights of UT 1989 Dec 23 and Dec 24 with the Faint Object Camera (FOCAM) at the prime focus of the 3.6m CFHT on Mauna Kea. An observing log with the details of the filter bands, exposure times, and measured seeing is given in Table 1. The detector was the "PHX1" CCD, a cosmetically excellent 516×516 pixel Ford CCD providing an image scale of 0″.275 pixel$^{-1}$ and an unvignetted field of view of $\sim$140″. Both nights were photometric with subarcsecond seeing; in round numbers the FWHMs of stellar images were $\sim$0″.6 for the red and green images, and $\sim$0″.8 for the blue images. Images were taken through eight narrowband filters: five emission-line bands and three continuum bands. The continuum-band filters were carefully selected to sample only line-free regions of the spectrum. These images provide templates for subtracting the stellar continuum component from the emission-band images, as well as mapping out the line-free continuum colors in the galaxy. White dwarf standard stars from the list of Oke (1974) were observed throughout both nights to provide an absolute flux calibration.

The raw images were bias and flat-field corrected following standard techniques with the Ohio State implementation of the Lick Observatory VISTA package. The emission-line and continuum images were registered spatially using the centroids of field star profiles on the [O III]$\lambda$5007 image to establish a common reference frame for all of the images. Centroids were measured using a marginal-sum technique, and since there was no field rotation, image registration required only a simple translation using a standard shift-and-rebin operation.

The continuum images were scaled to make templates for subtraction from the emission-line images to produce pure emission-line maps using the techniques described by Pogge (1992). Typical rms registration errors for frames were between ±0.03 and ±0.05 pixels (∼ ±8 − 14 mas). Uncertainty in the off-band continuum scaling is ∼ ±2%, reflecting the relative photometric precision of our standard star data. The final images were transformed onto an approximate absolute flux scale and corrected for atmospheric extinction using the standard Mauna Kea extinction curve. Further correction was made for Galactic reddening along the line of sight towards Mrk 573 by adopting $E(B-V) = 0.02$ derived from the tabulation of Burstein & Heiles (1984). The final data set consists of continuum-subtracted emission-line images of Mrk 573 in [O I], [O II], [O III], [S II], and H$\alpha$+[N II], and calibrated emission-free continuum images at effective rest-frame wavelengths of 3540, 5280, and 5995Å.

The flux calibration of the monochromatic images is complicated by changes in the laboratory filter bandpasses caused by their placement in the converging $f/4.1$ beam of the CFHT prime focus, and their use at low temperatures ($T \sim 0°C$). As a result, while our continuum flux calibrations are very repeatable and accurate, the monochromatic line flux calibrations require further refinement. The best procedure we have found is to perform a relative (rather than absolute) calibration of the emission-line fluxes using previous long-slit spectrophotometry. This is accomplished by synthesizing the slits on our emission-line images. The corrections to the monochromatic fluxes derived from the white-dwarf standard stars and laboratory filter transmission curves from this procedure are factors of order unity, but in extreme cases the relative intercalibration of the monochromatic line-flux images is affected at the 20 − 50% level. In the absence of detailed ray tracing of the filters to understand the precise bandpass distortion in the fast converging beam (e.g., as in Jacoby et al. 1989), this offers the best solution. The widths of our filters are significantly larger (50–75Å) than the observed systematic Doppler shifts in the extended emission regions in Mrk 573 (±200 km s$^{-1}$; TW92), so we are confident of our absolute line-ratio measurements at the ±15% level, while the relative line ratios within a particular map are precise to ±5%.

## 3. Results

### 3.1. Continuum

Figure 1a shows a contour map of the sum of the red (5995Å) and green (5280Å) continuum-band images. The lowest contour is at 0.1% of the peak nuclear flux, approximately 2$\sigma$ above the mean background sky level, and subsequent contours are spaced at roughly logarithmic intervals up to 90% of the peak flux level. Recall that the RC3 classification of this galaxy is RSAB(rs)0$^+$. Clearly visible in the contour map is the weak North-South bar and the faint outer stellar ring that gives this galaxy its "RSAB" classification. In the inner $5''$, the central isophotes give the appearance of a possible weak inner stellar bar oriented about East-West along $PA = 100°$, roughly perpendicular to the larger North-South bar (see Figure 1a). Figure 1b, on the same scale as Figure 1a, shows a logarithmic gray-scale representation of this image with contrast level chosen so as to enhance the inner ring and flocculent spiral arm-like features that appear in the starlight distribution, primarily defining the North and South ends of the weak bar. This is the feature which gives the galaxy its "(rs)" designation.

Figures 2a–c show the position angle, eccentricity and natural logarithm of the intensity along the major axis as a function of radius for the red continuum image, determined by fitting ellipses using standard azimuthal Fourier moment techniques. Within $4''\!.7$ the position angle is $\sim 98°$, while for $r > 5''\!.5$, the position angle rotates through $\sim 20°$ between about $170 - 190°$. The "transition region" between $5'' - 8''\!.5$ betrays the presence of a short N-S bar and weak spiral arms. (The behavior of the green continuum image is qualitatively similar except that the "transition region" is not as pronounced.) The eccentricity as shown in Figure 2b varies significantly with radius throughout the image. The isophotal contours are most circular at $5''$, and are most elongated at $9''$. If this latter distance represents the large-scale shape of the galaxy itself, and if the galaxy has a circular profile if viewed face-on, then the galaxy is inclined at an angle of $\sim 37°$ to the line of sight, tilted nearly along a N-S axis. The surface brightness profile shown in Figure 2c is not characteristic of a simple bulge or bulge-plus-disk, but a detailed surface brightness analysis is beyond the scope of this paper.

### 3.2. Emission-Line Images

Continuum-subtracted emission-line images of Mrk 573 in [O I], [O II], [O III], H$\alpha$+[N II], and [S II] emission are shown in Figures 3a–e as flux contour maps. For reference, the inner portions of the continuum contour map are reproduced in Figure 3f. The lowest contour in each case is roughly $2\sigma$ above the residual background level after sky and continuum subtraction, and subsequent contours are spaced at approximately logarithmic intervals up to 90% of the peak flux in the nucleus. We used two H$\alpha$ narrowband filters: the "H$\alpha_{\rm N}$" filter which has a 30Å bandpass that primarily transmits the H$\alpha$ emission line with a minor contribution from [N II] in the filter's wings, and the "H$\alpha$+[N II]" filter which has a 100Å bandpass that transmits the entire line blend. Only the H$\alpha_{\rm N}$ filter image is shown as the H$\alpha$+[N II] filter image shows the same basic structure albeit with lower contrast.

H II regions are seen most clearly in the H$\alpha_N$ image where they outline short, low pitch-angle spiral arms at a projected radius of $\simeq 8 - 9''$ from the nucleus. These arms do not appear to "originate" from the terminus of the stellar bar, but rather at a position angle near the symmetry axis of the ionization cone at $\sim 120°$, though there is certainly some confusion in this region emission from between H II regions in the spiral arms and extended emission associated with the active nucleus. This potential confusion is most clearly seen in the [O III] image where the H II regions are very faint (as expected for relatively cool, metal-rich H II regions in the interior of a galaxy) and the line emission is dominated by the complex extended nuclear emission regions (see Fig. 4). Perpendicular to the symmetry axis at a position angle of $34°$, the emission-line gas shows little spatial extension, consistent with previous imaging and long-slit observations. The overall distribution of circumnuclear emission-line gas shows the "bow-tie" (Haniff et al. 1988) or "bicone" (TW92) pattern reported for this galaxy. Moreover, these high-resolution data, especially the [O III] and H$\alpha$ images, reveal that the circumnuclear region is very richly structured down to the limit of our seeing ($0\rlap{.}''60 \pm 0\rlap{.}''05$ FWHM).

We note here that the symmetry axis defined by the kinematical model of TW92 shares the axis of symmetry of the extended [O III] emission-line gas; a position angle of $124 \pm 1°$. However, the position angle of the minor axis of the host galaxy measured from the outer isophotes ($R > 15''$) of our continuum images is significantly different, $\sim 101° \pm 1°$. For a simple solid-body rotation pattern, as suggested by TW92 or Whittle et al. (1988), one would have expected the morphological and kinematical minor axes to align. This difference could be attributed to the fact that the rotation axis of the ISM in some early-type galaxies can be different from the stellar rotation axes (e.g., Veilleux et al. 1993), or that the kinematics of this gas are in some sense decoupled from the gas in the disk of the galaxy itself. A detailed 2-D velocity field map obtained with either Fabry-Perot (e.g., Cecil, Bland, & Tully 1990) or Integral Field spectroscopy (e.g., Pecontal & Ferruit 1994) would be useful for investigating this matter further.

As is characteristic of most classical Seyfert galaxies, there is a very strong nuclear source that appears star-like in each of our images. We shall use the centroid of this nuclear source to define the "center" of the galaxy. To within our measurement uncertainties ($\sim \pm 0\rlap{.}''05$), the centroid of the nuclear source is the same as the centroid of the isophotes in the line-free continuum images, particularly the red image. Extending outward from the nucleus and roughly following the symmetry axis of the bicone are two emission-line intensity maxima located at $\sim 1\rlap{.}''72$ (SE) and $\sim 2\rlap{.}''25$ (NW) $\pm 0\rlap{.}''10$, respectively, from the center. These structures are not strictly collinear; the SE structure is centered along position angle $111 \pm 1°$, while the NW structure is centered along position angle $-63 \pm 1°$, a difference of $\sim 6°$, with the mean axis rotated $\sim -10°$ from the overall emission-line structure axis along 124°. Although the signal-to-noise (S/N) ratio varies considerably among the different emission-line images, it is clear that these two features have very extended arc-like or 'C' shapes reminiscent of a bowshock. This arc-like appearance is particularly evident in both of these structures in the [O I] emission-line image, while in our other emission-line images the NW region shows this morphology more clearly than the SE region. Overall, the two emission features are spatially resolved from the nuclear component, especially in the images of the low-ionization emission-line gas ([O I] and [S II]).

The [O III] and H$\alpha_N$ images have the greatest dynamic range of all of our images, and show additional emission-line structures at larger radii than those described above. Lying about 3″.30 SE and 4″.30 NW (±0″.10) from the nucleus (measured along PA=124° in the [O III] image) there are two extended, arc-like features. Rather than being positioned symmetrically with respect to the bicone symmetry axis (as is the case with the inner arc-like structures), these two regions are elongated nearly tangentially to this axis and rotated from the symmetry axis in a counter-clockwise direction, following the same sense of rotation as the spiral arms visible in the H$\alpha_N$ image. The effect (especially visible in the contour maps) is to lend a distinct S-shape to the distribution of emission-line gas within ∼6″ of the nucleus. In the absence of high spatial-resolution kinematic information it is impossible to determine precisely what is occurring here, although the sense of the change from roughly symmetric arcs to decidedly skewed arcs with increasing radius suggests a change from motions dominated by a nuclear outflow, to motions significantly modified by rotational shearing of the gas as it orbits about the center of the galaxy. A further clue to the origin of these nested structures is found in their relative locations. For both the inner and outer pairs of arcs, the NW arcs are located ∼ 30% farther from the nucleus in projected distance than their SE counterparts. The ratio of the projected radii of the outer peak to the inner peak on a given side, however, is the *same*, $(R_{out}/R_{in}) = 1.91$, to within our measurement uncertainties. We will discuss the implications of this below.

At the faintest isophotal levels in these [O III] and H$\alpha$ images, there are further emission-line complexes at a radial distance between about 7″ and 13″ that are nearly coincident with the spiral arms, yet aligned to within 2° of the axis passing through the inner arcs. Unlike the arc-like structures described above, these complexes have strong [O III] emission relative to H$\alpha$+[N II], and a very filamentary appearance. This is shown in Figure 4, where we have clipped the higher data values in the [O III] image and increased the contrast to show the faint outer structures. In this figure the [O III] flux contours are superimposed in white (with an arbitrary logarithmic spacing) to show the corresponding inner structures. Although kinematical information is lacking, these complexes could possibly represent gas clouds within the galaxy itself which, by virtue of lying inside the ionization cones, are photoionized by the central source, thus enabling them to maintain such a high level of ionization. The close coincidence between the axes of the inner arcs and the axes of these outer structures suggests that the escape axis of the ionizing photons is along $\sim 112°$, rather than along the overall inner isophotal symmetry axis of $\sim 124°$ found from the bright emission-line complex within $6''$ of the nucleus that is seen in the maps of the different emission lines.

### 3.3. Physical Conditions and Line Ratios

In the bright [O III]$\lambda 5007$ emission-line, the extended regions outside the nucleus account for $\sim 80\%$ of the total [O III] luminosity in this galaxy, with only $\sim 20\%$ coming from the unresolved, "classical" narrow-line region (NLR). Fully 90% of the total [O III] emission (including the nucleus) comes from within a radius of $4''\!.8$ from the nucleus ($\sim 1.55 h_{75}^{-1}$ kpc), encompassing the two sets of arc-like emission-line regions that bracket the nucleus.

Some of the physical parameters in the extended emission-line gas may be estimated by making a few reasonable assumptions. First, consider the innermost arc-like emission-line regions (hereafter, simply "arcs") which contribute approximately 9% (NW) and 17% (SE) of the total [O III] emission in the circumnuclear regions. The arcs are marginally resolved and are situated at a projected distance of $\sim 0.6 h_{75}^{-1}$ kpc from the nucleus of the galaxy.

If the ratio of the ionization parameter in the NLR [O III] gas to the ionization parameter in the arc gas is $\gamma$, then to an order of magnitude we can write:

$$n_{arc} \approx \gamma\, n_{nlr}\, \left(\frac{R_{nlr}}{R_{arc}}\right)^2 \approx 2 \times 10^3\, \gamma\, \text{cm}^{-3}, \tag{1}$$

using $R_{nlr} = 30 h_{75}^{-1}$ pc and adopting $n_{nlr} = 7 \times 10^5$ cm$^{-3}$, the critical density of [O III]$\lambda 5007$ because Mrk 573 displays a reasonable line width-critical density correlation (De Robertis & Osterbrock 1986). For the more diffuse and extended emission-line gas, the number density is thus expected to be about an order of magnitude less than that inferred for the NLR.

The volume filling factor of the extended gas may also be related to the filling factor in the NLR, $f_{nlr}$. Consider first the arcs. The observed [O III] intensity $I_{arc}$ is,

$$I_{arc} \propto n_{arc}^2\, f_{arc}\, \frac{4}{3}\pi R_{arc}^3\, (1 - \cos\theta)/2, \tag{2}$$

where $\theta$ is the full opening-angle of the ionization cone. The intensity from the classical (spherically symmetric) NLR is

$$I_{nlr} \propto n_{nlr}^2\, f_{nlr}\, \frac{4}{3}\pi R_{nlr}^3. \tag{3}$$

Using the parameters measured above, and assuming that (a) there is no significant attenuation of the line radiation in either region, (b) the regions have similar chemical compositions, and (c) photoionization in both regions is due to the central continuum source, then

$$f_{arc} \approx \frac{5}{\gamma^2} \, f_{nlr} \left(\frac{R_{arc}}{R_{nlr}}\right) \approx \frac{10^2}{\gamma^2} \, f_{nlr}. \tag{4}$$

The analysis of TW92 indicated that $\gamma \approx 5 - 10$, suggesting that the filling factor in the arcs is roughly the same as that in the NLR and the diffuse line-emitting regions. In this case, the gas in the arcs would have a density of $\sim 10^4$ cm$^{-3}$, of the same order seen in dense Galactic nebulae (such as Orion, e.g., Osterbrock 1989). If instead $\gamma \sim 1$, then this would require that $f_{nlr} << 1$ in order for $f_{arc}$ to have a reasonable value.

Figures 5a–c show contour maps of the emission-line ratios [O I]/[O III], [O II]/[O III], and [O III]/H$\alpha$+[N II] (actually [O III]/H$\alpha_{\rm N}$), derived from the images shown in Figure 3. For comparison, Figure 5d reproduces the [O III] emission-line flux contour map of Figure 3a. No reddening correction has been applied to these data, apart from the Galactic reddening correction noted above. Internal reddening for the nuclear spectrum has been derived from a variety of methods (e.g., Koski 1978; Malkan 1983; MacAlpine 1988; TW92) to be $E(B - V) \approx 0.30$. The Balmer decrement (H$\alpha$/H$\beta$) has been measured for some of the extended line gas by TW92 who find that the observed H$\alpha$/H$\beta$ intensity ratios vary between 3 in the NW, to a maximum of 5, some 2″–2″.5 to the SE. Thus the extinction may be patchy and cannot be dealt with in a simple way, though there is no evidence from the continuum color maps (Pogge & De Robertis 1993) to suggest that the fine structure we see is simply an artifact of extinction, as the coincidences in geometry noted above and the trends in line ratios described below show.

Figures 6a and 6b show the emission-line ratios of [O I]/[O III] and [O II]/[O III] as a function of projected radius from the nucleus measured along position angle 124°. On each spatial cut we have marked the locations of the centroids of the off-nuclear 6 cm radio continuum lobes measured from the original UW84 radio map (kindly provided by J. Ulvestad and A. Wilson). As is seen in the contour maps (Figure 5), the inner emission-line arcs stand out as regions of lower relative ionization than their surroundings, with a peak at a larger radius than the radio continuum emission knots. We will discuss the relationship between the optical and radio data further in the next section (§3.4.).

We can also discern the effect of differential reddening between the SE and NW sides noted by TW92 in our [O II]/[O III] ratio map. This can be seen most clearly in Figure 7 where we reflect the spatial cuts in Figure 6b about the nucleus, and plot the SE side as filled circles and the NW side as open circles. The [O II]/[O III] ratios measured on the SE side of the nucleus are systematically smaller than those seen on the NW side at the same radius. This, coupled with the greater extension of bright structure at large radii seen towards the NW on the [O III] images especially (Figure 4), suggests that the NW side is emerging towards us slightly out of the plane of the host galaxy, while the SE side is receding, and subject to slightly higher extinction from the disk of the host galaxy. This effect has been seen in other Seyferts with ionization cones, in particular in NGC 5728, where both ground-based and HST images show the receding ionization cone disappearing behind a circumnuclear ring of H II regions (Pogge 1989; Wilson et al. 1993), to then barely reappear just beyond the radius of the H II region ring. This also agrees with the kinematical information presented in TW92 and Whittle et al. (1988).

The forbidden oxygen line ratios are particularly sensitive to the (local) ionization parameter, $U$, and in the region of parameter space of interest to us, the [O I]$\lambda 6300$/[O III]$\lambda 5007$ and [O II]$\lambda 3727$/[O III]$\lambda 5007$ line ratios vary monotonically with $U$. Figure 8 illustrates the oxygen emission-line parameter space occupied by both published observational data and theoretical models. The utility of one or more of these line ratios as a diagnostic of the degree of ionization in the line-emitting gas has been long recognized, and was an important feature in the classification scheme originally described by Baldwin, Terlevich, & Phillips (1981). More recent work, including modifications of this scheme by Veilleux & Osterbrock (1987), has concentrated on using those line ratios with small wavelength baselines to minimize the effects of extinction. Despite the reddening sensitivity of the oxygen line ratios, Figure 8 shows that they nonetheless provide an excellent diagnostic of the ionization state of the gas and can allow us to distinguish between various possible ionization mechanisms.

The dotted-line rectangles in Figure 8 show the regions of the parameter space spanned by published observational data for H II Regions, Seyferts, and Liners. The rectangle labeled "H II Regions" is derived from Evans & Dopita (1985), who used the spectral data of McCall et al. (1985) with measurable [O I]$\lambda 6300$ emission. The area depicted includes more than 85% of the H II regions reported in that study. The theoretical models presented by Evans & Dopita (1985) span a similar region in this diagram. The rectangle labeled "Seyfert 1, 2" is based on the spectrophotometric data used by Ferland & Netzer (1983; hereafter FN83), and includes Seyfert 2 galaxies and "broad-line objects." The rectangle labeled "Liners" is also based on the range of line ratios observed in the published spectrophotometry considered by FN83.

The solid line is the photoionization model computed by FN83 which assumes a power-law ionizing continuum with a slope $\alpha = -1.5$ and no short-wavelength cutoff, a constant gas pressure, a density of $10^3$ cm$^{-3}$, and no extinction. The open symbols plotted on this curve denote specific values for $U$: an open triangle, open square, and hexagonal star represent $U = 10^{-2}$, $10^{-3}$, and $10^{-4}$, respectively. This specific model sequence reproduced both the Seyfert (high ionization) and Liner (lower ionization) observations reasonably well. (It is important to note FN83 found that these oxygen ratios are insensitive to the metallicity of the gas for values between 0.1 and 1 times the solar abundance.) The dashed curves are the integrated photoionization models of Binette (1985), with power-law slopes $\alpha = -1$, $-2$, and $-3$ ($\alpha = -2$ provided the most consistent matches to the observed Seyfert and Liner line ratios). The open symbols plotted on the curves are the same as for the FN83 data, with an additional open circle plotted to represent $U = 10^{-5}$.

The rectangle labeled "Shock Models" and drawn with short-dashed lines is the region of the diagnostic plane spanned by the high-velocity shock models of Binette, Dopita, & Tuohy (1985). The "Liners" parameter space is almost completely contained by these shock models. More recent shock models of Sutherland et al. (1993) occupy the gray-shaded rectangle located within the "Seyfert 1, 2" box.

Oxygen line ratios for the nucleus and the two inner emission-line arcs in Mrk 573 are plotted in Figure 8 as filled points with error bars. A reddening vector for $E(B - V) = 0.30$ mag, the estimate of the extinction in the nucleus from TW92, is also plotted. These data are consistent with "Seyfert-like" ratios, indicative of photoionization by a hard continuum. With the exception of a couple of points with large uncertainties, plotting all of the data along the spatial cuts in Figures 6a and 6b fills the portion of the "Seyferts 1,2" box to the left of points for the inner SE and NW arcs. It appears then that the ionization parameter drops sharply in the arcs and rises in the emission-line gas further out, subject to the usual extinction caveat.

Based on Figures 6–8, it is apparent that at large radii the ionization parameter rises, except when passing through the inner arc-like structures where it falls. The trend of a slightly increasing ionization parameter with radius well outside the nucleus was also noted by TW92 based on their long-slit spectra. Their favored explanation suffices for our data as well: the trend is due primarily to the density of the extended ionized gas decreasing more rapidly with radius than $r^{-2}$, leading to a rising ionization parameter. This explains the large-scale trends, and is consistent with the relatively rapid fading of the extended emission-line regions with larger radius. Impressed on this, however, our higher-resolution imaging data has resolved finer-scale structure that we interpret as the result of an increased density in the arcs. The arcs are barely resolved in our data, and may in fact be dominated by unresolved structure. A possibility we will discuss below is that the arcs are edge-brightened radiative bowshocks where the outflowing radio plasma plows into the circumnuclear ISM of the galaxy. The higher density inferred from the ionization parameter arguments would then be a consequence of compression in the shock zone.

Interestingly, the [O III]/Hα+[N II] ratio map for this galaxy shown in Figure 5c, the more traditional "ionization" map, exhibits little evidence of coherent structure. That this might be the case was hinted at by the lower spatial-resolution ionization map of TW92. The map in Figure 5c is very lumpy, reflecting the fine structure visible in both the [O III] and Hα$_N$ filter images. The classical ionization cone is hardly discernible any longer in this map compared with what is seen in our other line ratio maps (Figures 5a and 5b). From Figure 8, it appears that a more robust relative ionization indicator is provided by ratios of lines from different ionization stages of the same atom such as oxygen.

As a general remark, we have noticed that as higher spatial resolution images have become available from ground-based and HST observations, often the traditional ionization maps retain very little of their familiar (bi)conical morphology. The impression is that with improved spatial resolution the cones "break up" into finer and finer structures and become less well-defined. (This makes extended emission-line regions excellent candidates for study with emerging adaptive optics techniques.) On larger scales, one still sees the general shape of the ionization cones, even when there is considerable small-scale structure (e.g., Macchetto et al. 1994). This is particularly evident in the oxygen emission-line maps in Figures 5a and 5b. Thus, Mrk 573 appears to represent a case in which one cannot easily distinguish between an extranuclear morphology determined by the distribution of ionizing photons (i.e., a "pure" ionization cone), and a morphology determined by the distribution of emission-line gas.

### 3.4. Radio Data

Figure 9a shows the VLA 6 cm radio data of UW84 superposed upon a gray-scale [O III] image. Both the radio and optical data have comparable spatial resolutions (∼0$''$.6 and 0$''$.75 for the optical and radio maps, respectively). This figure demonstrates the general alignment of the radio and optical emission axes, as well as the substantially greater radial extent of the optical emission. The bright peaks of [O III] emission appear in somewhat lower contrast compared to the general diffuse [O III] emission component, so to better emphasize the relation between the radio and optical emission, we show in Figure 9b the radio flux contours superimposed on a version of the [O III] image that has been processed by subtracting a digital unsharp mask. The mask was generated by smoothing the original [O III] image with a uniform-weight $5 \times 5$ pixel "boxcar" kernel. This processing serves to enhance the spatially sharp arc-like features seen in all of the emission-line bands, and clarifies the point made previously that the off-nuclear radio emission sources lie just inside the first set of arc-like emission-line structures. In particular, note how the inner NW [O III] emission arc seems to bracket the brightest part of the NW radio lobe. The SW [O III] emission peak has a less distinct arc shape in this figure (it can be seen better in the [O I] image, Figure 3c), and the SW [O III] peak appears to be enveloped by the radio flux contours. The optical emission and radio nuclei precisely coincide on these images. This general displacement between the fine structure in the emission-line maps and the locations of the brightest off-nuclear 6 cm radio lobes is seen in the [O I]/[O III] line-ratio cut along the symmetry axis in Figure 6a. Note that in this plot the region of lowest effective ionization parameter (highest [O I]/[O III] ratio) is at larger projected angular distance than either of the radio peaks; $1''.95$ compared to $1''.60$ towards the SE, and $1''.96$ compared to $1''.25$ towards the NW for line ratio and radio continuum maxima, respectively. This is equivalent to a projected angular displacement of $0''.35$ and $0''.71$, SE and NW, respectively, corresponding to a projected radial displacements of 116 and $234 h_{75}^{-1}$ pc, respectively.

Finally, we point out an important correspondence between the lowest radio flux contours and the general shape of the inner pair of arc-like emission clouds. This can be seen in both panels of Figure 9 where the lowest radio flux contours show a twisting in the clockwise direction, following the general clockwise twist of the optical emission arcs. Indeed, in Figure 8b, the lowest radio flux contours appear to envelope the [O III] emission arcs, while the essentially unresolved bright cores of the radio lobes are nestled inside the arcs (more to the NW than the SE as discussed above). This striking similarity in general behavior between the optical and radio gas is further compelling evidence that the two are intimately related to each other.

## 4. Discussion

There is a strong tendency for the axes of the conical emission-line regions in some Seyferts to be very closely aligned with the radio continuum axes (Wilson & Tsvetanov 1994), suggesting a physical relationship between them. Less clear, however, is the detailed spatial correlation, if any, between the brightest radio plasma knots and the optical emission-line regions. Morphologically, a good example of a direct spatial correlation between optical line-emitting clouds and radio plasma lobes is found in NGC 5929 (Bower et al. 1994), where HST imaging shows no spatial displacement between the brightest radio and optical emitting regions. More convincing evidence of a direct physical interaction is the apparent kinematic relationship that has been shown by the long-slit data of Whittle et al. (1988), and clearly demonstrated in the best-resolved case thus far studied, NGC 1068, where discrete kinematic substructures in the line-emitting gas can be directly associated with the jet and the bowshock components of the radio continuum structure (Cecil, Bland, & Tully 1990).

– 15 –In Mrk 573 we can cite two pieces of evidence in favor of a similar direct radio-optical interaction: the long-slit kinematics of Whittle et al. (1988) and the morphology of our high-resolution emission-line images. As noted previously, at the location of the radio lobes, Whittle et al. see kinematic substructure in the [O III] line emitting gas, in the form of a complex, multi-component emission-line profile. While not as clear a demonstration as might be provided by a full 2-D velocity field from Fabry-Perot or Integral-Field spectroscopy (where one can attempt to exploit spatial correlations to decompose the gas motions, e.g., as Cecil et al. 1990 have done so effectively in NGC 1068), it is nonetheless highly suggestive.

In our high-resolution emission-line images, the highest surface-brightness ionized gas regions located just beyond the active nucleus appear to assume a shape characteristic of a bowshock. This is visible in all the emission lines, but the contrast between these regions and the general diffuse emission in the ionization cone proper is highest in the [O III], [O I], and [S II] images. The NE radio lobe of NGC 1068 has a similar bowshock appearance at 2 and 6 cm (WU87). Indeed WU87 consider a variety of models for this feature, and conclude that a radiative bowshock is indeed a reasonable possibility. The distinction here is that in Mrk 573, it is the *optical* emission which assumes this shape and not the radio emission.

The lack of an edge-brightened radio bowshock does not appear to be simply a problem of lower resolution and/or dynamic range, at least not entirely. In Figure 10 we simulate the effect of observing NGC 1068 at the distance of Mrk 573 ($\sim 67\, h_{75}^{-1}$ Mpc) by smoothing and resampling the WU87 6 cm radio map of NGC 1068 to the same effective beam and "pixel" scale as the Mrk 573 map, the spectacular edge-brightened radio bowshock in the NE lobe of NGC 1068 is almost unrecognizable: it looks like a classic three component linear radio source of roughly similar angular size, although some edge-brightening is still evident. For reference, we show in Figure 10 (a) the original NGC 1068 radio map of WU87, (b) our synthetic view of the same map at the distance and beam size of the UW84 map of Mrk 573, and (c) the UW84 Mrk 573 map side-by-side for comparison.

A possible explanation of the NGC 1068 radio bowshock is that we are seeing the compression of the ambient magnetic field and cosmic rays in the bowshock region. A likely source of these particles is NGC 1068's very powerful circumnuclear starburst. Since we do not see a similar starburst in Mrk 573, this population of energetic particles may be absent, reducing the synchrotron emissivity in the bowshock region (which WU87 estimate to be enhanced by about a factor of $\sim 10^4$ in NGC 1068 bowshock) below the detection limit of the present radio maps. Thus the extended radio power and morphology are dominated by the jet component, not the bowshock component. This offers a possible explanation for why the centroid of the radio plasma knots lies behind the apparent emission-line bowshock in Mrk 573. This is not to say, however, that there is no radio flux associated with the arcs. As can be seen in Figure 9b, there is low-level radio emission from the arcs that follows the same sense and magnitude of twisting in position angle, with the lowest radio flux contours appearing to envelope the emission-line arcs and following their outer isophotes.

There are a number of models that have been used to explain the source of radio emission in Seyfert galaxies. Jet-induced star formation and subsequent supernovae is one possibility. But because extranuclear star formation does not appear to contribute significantly to the photoionization of the gas — see §3.3 — we will consider only the two mechanisms described by WU87. That is, either the radio emission is due to a continuous supply of material from the nucleus in the form of a jet directed along the radio symmetry axis, or to the emission of a discrete "plasmoid" from the nucleus. (It is also possible that the jet precesses, thereby changing its apparent position angle with time.) It could be imagined that either of these models might lead to a "snow-plowing" of the interstellar medium (ISM) and hence to the production of a geometry similar to that which is observed in the optical line-emitting gas. Apart from the general appearance and the suggestive nature of the kinematical data, we cannot make a definitive case for a bowshock in this case. But *if* the radiative bowshock model of WU87 is appropriate for Mrk 573, then it might be expected that some of the physical parameters inferred for the NE lobe in NGC 1068 might also apply to the lobes in Mrk 573. The metric distance from the central source is certainly of the same order as for the lobe in NGC 1068 ($\sim 130\,h_{75}^{-1}$ pc in Mrk 573 as compared to $\sim 105\,h_{75}^{-1}$ pc in NGC 1068, measured from the nucleus to the bowshock in the latter).

The mass of gas swept up by the shock caused by the radio jet in NGC 1068 is between $10^{5-8} M_\odot$, depending on the pre-shock densities in the ISM. While this spans a large range in mass, the mass of ionized gas inferred from the [O III] emission in TW92 are within these limits. The bowshock in NGC 1068 is traveling radially with a velocity between $10^{2-3}$ km s$^{-1}$. If the shock in Mrk 573 travels with this same speed, then the "age" of the shock would be $\sim 10^{5-6}$ yr. (Because of the curvature of the optical lobes and the large ratio of length-to-width at a fixed isophotal level, the jet must be approximately in the plane of the sky. This is consistent with the inferred inclination angle of the disk as described in §3.1.) The propagation of the shock would be expected to generate large (turbulent) velocities in the tenuous gas downstream, i.e., between the bowshock and the nucleus. WU87 suggest that some and perhaps most the extreme velocities associated with the emission-line widths in some Seyferts could be linked to the effects of the radio jets (see also Whittle 1992a). Long-slit observations by Whittle et al. 1988 do show weak, but extensive, wings in the [O III] profiles. The largest blue wing is indeed found approximately at the outer edge of the NW optical feature, but the largest red wing occurs in the second SE optical emission maximum, not the first. This inconsistency is not that distressing, as these data were obtained for only a single position angle, and have lower spatial resolution than our imaging data, so the match is poor. The fact that they do see kinematic substructure in these regions makes a detailed 2-D kinematical study of this galaxy reasonable. Without these data, our evidence of a bowshock in this region must remain somewhat circumstantial.

While a radiative bowshock might be responsible for *shaping* the emission-line gas into an arc, we cannot clearly assess the degree to which the shocks also contribute to the ionization of the gas. The high-velocity shock models of Binette et al. (1985) illustrated in our diagnostic diagram in Figure 8 clearly do not reproduce the observed oxygen emission-line ratios in Mrk 573. More recent models by Sutherland et al. (1993), however, have shown that fast autoionizing shocks can produce Seyfert-like emission-line ratios under the right conditions. This shows that our simple diagnostic diagram, although useful, is incapable of unambiguously establishing the underlying ionization mechanism in all cases. The morphology of the inner emission arcs is reminiscent of the strong shocks described by Sutherland et al. (1993), but unlike the situate described by them these regions are embedded in ionized gas extending well beyond the radio plasma jet and are presumably ionized by radiation from the active nucleus. This suggests a composite situation in which ionization in the arcs is very likely provided by both mechanisms. Unfortunately, we have insufficient data to assess the relative importance of these competing emission mechanisms. While the Sutherland et al. (1993) models disagree with our observations, we note that their models were tailored to apply to a specific emission-line region (the filaments in Centaurus A), and so we cannot rule out this class of models a priori because of failure to agree in detail with data for our particular emission-line regions. Further observational and theoretical work is required to explore this possibility further, particularly spectrophotometry in the satellite UV with HST (where Sutherland et al. predict that lines like C IV$\lambda$1549Å appear to provide good discrimination between autoionizing shocks and power-law photoionization), and grids of both pure shock and hybrid photoionization and shock models spanning more of the relevant parameter space to provide a sound basis for comparison with the observations.

Finally, we wish to comment speculatively on a curious feature of our emission-line images that perhaps bears on the evolution of the circumnuclear environment in this galaxy. We see two pairs of arc-like emission-line maxima that lie along the same approximate axis of symmetry on either side of the active nucleus, and in which the ratio of the radii of the outer and inner arcs is the same ($R_{out}/R_{in} = 1.91$) on both sides (as noted in §3.2). This could be taken to indicate that if a nuclear outflow is responsible for organizing the optical gas, it might be an episodic rather than steady-state outflow. That is, the outflow could be "on" for a period, and then "off." The plasmoid models possess this feature naturally. The radio jet models would require the supply from the central source to be shut off or temporarily interrupted for a period. This is consistent with our observation that at least in the emission lines, the arcs appear to be almost completely separated from the nucleus. In the "off" period the optical gas will continue to move radially outward because of its bulk motion. Indeed, it can be shown that "ram pressure" is sufficient to slow significantly $\sim 10^6$ solar masses of radially moving material with an initial velocity of a few hundred km s$^{-1}$ within $\sim 10^6$ yr. In the same way, the material would also acquire a rotational velocity component during its encounter with the ambient ISM which is partaking in the general rotation of the disk of the galaxy. We note that the sense of the observed rotation in position angle between the inner and outer arcs is in the right sense expected from the observed velocity field (e.g., Whittle et al. 1988; TW92) if the outer (and presumably "older") arcs had been rotationally "sheared" by the ambient ISM.

## 5. Conclusions

Our high-resolution, optical emission-line and continuum images have revealed the tremendous wealth of structure in the extended narrow-line region of Markarian 573. The most striking features we have found are two pairs of arc-like emission-line structures that bracket the nucleus. The innermost pair lies within the central 2″ of the nucleus, partially enveloping the two lobes of extended radio continuum emission which lie on either side of the nuclear radio source. In these arcs we see a rapid change in the ionization state of the gas (revealed by our line-ratio maps), which when coupled with the morphology strongly suggests that these structures are kiloparsec-scale bowshocks that have been driven into the ISM of the host galaxy by the radio plasma ejected from the active nucleus. The outermost arcs are elongated perpendicular to the symmetry axis defined by the radio emission, and have the appearance of being sheared tangentially in the direction of rotation of the galaxy. Taken together, the alignment and relative positions of these four emission features suggest that whatever is responsible for the geometry of the emission-line gas may be episodic in nature.

The morphological and spectrophotometric results for Mrk 573 emphasize a growing realization that we can no longer model these regions using simple photoionization models alone. Nor can we take the suggestions of a possibly strong mechanical interaction with the radio jet in this and in a growing number of Seyfert galaxies (especially with new HST data from a variety of collaborations) as evidence that only shocks are important. That there exists a considerable amount of ionized gas *beyond* the regions of apparent interaction with the radio emitting material suggests that we must consider hybrid models in which shocks are driven into an ISM that has been pre-ionized by the nonthermal nuclear source. A challenge for future observational work will be to clarify the nature of this interaction with higher spatial resolution data (both at radio wavelengths and with the HST), and with detailed kinematic and spectrophotometric observations to try to disentangle the different regions by their motions and physical conditions. On the theoretical side, there is the challenge to devise new emission-line diagnostics that will have the power to distinguish between the two ionization mechanisms, both to assess the relative importance of each and to confirm physically the visual evidence for shocks.

The authors would like to thank M. McCall for the use of his interference filters, and to L. Binette, P. Osmer, B. Peterson, D. Axon, and J. Morse for insightful discussions. We are deeply indebted to J. Ulvestad & A. Wilson for providing the original radio continuum maps, and J. Mulchaey for converting them into FITS format. The referee, C. Haniff, gave the paper a remarkably thorough and critical review that substantially improved the final results. MMDR acknowledges financial support from the Natural Science and Engineering Research Council of Canada. RWP is indebted to the OSU Astronomy Dept., and received partial financial support from a scholarship from the Northern California Chapter of the ARCS Foundation. This research has made use of the NASA/IPAC Extragalactic Database (NED) which is operated by the Jet Propulsion Laboratory, Caltech, under contract with the National Aeronautics and Space Administration, and NASA/ESA Hubble Space Telescope images obtained from the data archive at the Space Telescope Science Institute. STScI is operated by the Association of Universities for Research in Astronomy, Inc., under NASA contract NAS5-26555.

### Note Added in Manuscript

After the revised version of this paper was resubmitted to the journal, a set of Post-COSTAR Hubble Space Telescope Faint Object Camera (FOC) narrowband images of Markarian 573 became publically available in the HST Archives. Figure 11 shows a segment of the F502M filter image (ID number X2580503t) showing the active nucleus and extended [O III] emission-line clouds. Due to mis-centering of the galaxy on the FOC, only the nucleus and NW extended regions are visible. The F502M filter includes both stellar continuum and the [O III]5007Å emission line at the redshift of Mrk 573, and for this image we have not attempted to subtract the stellar continuum image that is part of this data set. On this image we have superimposed the 6 cm radio flux contours, using the same contour levels as shown in Figure 9a and 9b. This FOC image, with a spatial resolution of about $0''\!.1$ FWHM, beautifully confirms our CFHT observations that the inner [O III]-bright region located to the NW of the nucleus has a distinctly arc-like morphology that wraps around the peak of the 6 cm radio continuum emission. The resemblance between the shape of the NW arc and the bowshocks seen at the heads of some Galactic Herbig-Haro object jets is remarkable.

---

## Figure Captions

Figure 1. *a*) Flux contour map of the sum of the red and green continuum images of Mrk 573. The levels are logarithmically spaced, with the lowest approximately 0.1% of the peak nuclear flux and $2\sigma$ above the mean sky background. North is up and east is to the left. *b*) A gray-scale representation of the same image with the contrast adjusted to emphasize the inner ring and flocculent spiral arms.

Figure 2. Surface photometry of the red continuum image of Mrk 573: *a*) (top) major-axis position angle as a function of projected semi-major axis radius in arcseconds (note the sharp discontinuity at $\sim 4.5 - 5$ arcseconds), *b*) (middle) the contour eccentricity, and *c* (bottom) the natural logarithm of the azimuthally averaged surface brightness.

Figure 3. Contour diagrams for pure emission-line images of [O III], H$\alpha_N$, [O I], [S II], and [O II]. The lowest contour in each case is roughly $2\sigma$ above the residual background level after sky and continuum subtraction, and subsequent contours are spaced at approximately logarithmic intervals up to 90% of the peak flux per pixel as follows: (a) [O III]: (0.1, 0.2, 0.3, 0.5, 0.7, 1, 1.5, 2, 3, 5, 7, 10, 15, 20, 30, 50, 70, 90); (b) H$\alpha_N$: (0.15, 0.2, 0.3, $\cdots$, 90); (c) [O I]: (1, 1.5, 2, $\cdots$, 90); (d) [S II]: (0.7, 1, 1.5, $\cdots$, 90); (e) [O II]: (2, 3, 5, $\cdots$, 90). Note the extended emission-line material which is roughly biconical in shape (with a symmetry axis $\sim 124°$) but with rather complex small-scale structure. There is little extended emission perpendicular to the symmetry axis at the nucleus. Stubby spiral arms are evident in the H$\alpha_N$ image. 3*f*) shows a continuum contour diagram on the same scale as 3a–e for comparison. Contour intervals relative to the peak nuclear flux are the same as for Figure 3a. Note the absence of any continuum features along the biconical symmetry axis.

Figure 4. Logarithmically spaced [O III] flux contours displayed on a gray-scale image of the same with the brightest regions clipped to enhance the contrast in the faintest emission-line structures. Note the extensive filamentary emission-line structures oriented roughly along the biconical symmetry axis.

Figure 5. Line ratio maps. *a*) (upper left) Line-ratio map of [O I]/[O III]. Contours are superimposed on a grayscale representation of the line ratio (darker corresponds to larger line ratios), with values of [O I]/[O III]=0.01–0.05 in steps of 0.01, plotted as white lines for [O I]/[O III]$\geq 0.04$. This ratio is particularly sensitive to the ionization parameter. *b*) (upper right) Line-ratio map of [O II]/[O III]. Contours run from 0.1–0.3 in steps of 0.02, and (0.25,0.3) at the high end. Contours are plotted as white lines for [O II]/[O III]$\geq 0.2$. *c*) (lower left) Line-ratio map of [O III]/H$\alpha_N$. Contours run from [O III]/H$\alpha_N$=1.0–4.0 in steps of 0.25, with contours plotted a white lines for [O III]/H$\alpha_N \geq 2.0$. At this spatial resolution, this "ionization map" is very complex without evidence for a coherent large-scale ionization

cone structure. *d*) (lower right) is a repeat of the [O III] flux map (Figure 3a) to provide a relative scale for panels $5a - c$. In both panels *c* and *d*, the field of view is twice that in panels *a* and *b*.

Figure 6. *a*) NW–SE spatial cut along the biconical symmetry axis and through the nucleus of the [O I]/[O III] line ratio image. The ionization parameter increases as this ratio decreases. The uncertainties for each measurement are included as error bars. The position of the centroids of the NW and SE radio continuum lobes are indicated by short-dashed lines. Note that the radio centroids lie just *inside* the peaks in the line ratio spatial cuts. *b*) Same as in *a* but for [O II]/[O III].

Figure 7. Same as in 6*b* but reflected about the same side of the nucleus. The NW side ratios (open circles) are systematically higher than the SE side (filled circles) suggesting that the SE side may be located on the "far side" of the galaxy, while the NW side is on the "near side," consistent with the suggestions of previous kinematical data.

Figure 8. Oxygen emission-line diagnostic diagram. See the text for details.

Figure 9. *a*) VLA 6 cm radio map superimposed on our gray-scale [O III] image. The data have essentially the same spatial resolution. Note the general alignment of the radio lobes and the extended optical line emission. *b*) The same as in *a* but now the radio flux contours are superimposed on an unsharp masked version of the [O III] image to emphasize the spatially sharp arc-like features that bracket the active nucleus. As in Figure 6, the radio lobes lie inside the emission-line arcs. The secondary emission-line "tangential" features are also evident in this plot, well outside the radio emission.

Figure 10. *a*) The original Wilson & Ulvestad (1987) 6 cm radio map of NGC 1068 at full resolution. *b*) This same map as it would appear if NGC 1068 were at the distance of Mrk 573 and observed with the same resolution. For reference, panel (*c*) shows the Ulvestad & Wilson (1984) VLA map shown for reference at the same angular scale as panel *b*.

Figure 11. VLA 6 cm radio flux map superimposed on a gray-scale representation of an HST post-COSTAR FOC [O III] emission-band image (F502M filter) of the central regions of Mrk 573. Radio flux contours are plotted the same as in Figures 9a and 9b.

Table 1. Journal of Observations

| Band | Filter Parameters | | UT Date | Exposure | Seeing[a] |
|---|---|---|---|---|---|
| | $\lambda_C$ | FWHM | | | |
| [O II]$\lambda$3727 | 3765Å | 75Å | 1989 Dec 23 | 1800s | 0.″76 |
| [O III]$\lambda$5007 | 5100Å | 96Å | 1989 Dec 23 | 600s | 0.″55 |
| [O I]$\lambda$6300 | 6405Å | 50Å | 1989 Dec 24 | 600s | 0.″57 |
| H$\alpha$ Narrow | 6675Å | 39Å | 1989 Dec 23 | 900s | 0.″63 |
| | | | 1989 Dec 24 | 600s | 0.″56 |
| H$\alpha$+[N II] | 6700Å | 100Å | 1989 Dec 24 | 720s | 0.″58 |
| [S II]$\lambda$6716+6731 | 6830Å | 75Å | 1989 Dec 24 | 900s | 0.″64 |
| UV Continuum | 3600Å | 200Å | 1989 Dec 23 | 1800s | 0.″76 |
| Green Continuum | 5370Å | 192Å | 1989 Dec 23 | 600s | 0.″56 |
| Red Continuum | 6100Å | 198Å | 1989 Dec 24 | 300s | 0.″57 |

[a]Measured FWHM of field stars.